# Cyclinac Medical Accelerators Using Pulsed $C^{6+}/H_2^+$ Ion Sources

A. Garonna[a,b*], U. Amaldi[a], R. Bonomi[a,c], D. Campo[a], A. Degiovanni[a,b], M. Garlasché[a], I. Mondino[a], V. Rizzoglio[a,d] and S. Verdú Andrés[a,e]

[a] *TERA Foundation, Geneva, Switzerland*
[b] *Ecole Polytechnique Fédérale de Lausanne EPFL, Switzerland*
[c] *Politecnico di Torino, Italy*
[d] *Università degli Studi di Torino, Italy*
[e] *Instituto de Física Corpuscular IFIC (CSIC-UVEG) , Valencia, Spain*
   E-mail: Adriano.Garonna@cern.ch

ABSTRACT: Charged particle therapy, or so-called hadrontherapy, is developing very rapidly. There is large pressure on the scientific community to deliver dedicated accelerators, providing the best possible treatment modalities at the lowest cost.

In this context, the Italian research Foundation TERA is developing fast-cycling accelerators, dubbed 'cyclinacs'. These are a combination of a cyclotron (accelerating ions to a fixed initial energy) followed by a high gradient linac boosting the ions energy up to the maximum needed for medical therapy. The linac is powered by many independently controlled klystrons to vary the beam energy from one pulse to the next. This accelerator is best suited to treat moving organs with a 4D multi-painting spot scanning technique.

A dual proton/carbon ion cyclinac is here presented. It consists of an Electron Beam Ion Source, a superconducting isochronous cyclotron and a high-gradient linac. All these machines are pulsed at high repetition rate (100-400 Hz). The source should deliver both $C^{6+}$ and $H_2^+$ ions in short pulses (1.5 μs flat-top) and with sufficient intensity (at least $10^8$ fully stripped carbon ions at 300 Hz). The cyclotron accelerates the ions to 120 MeV/u. It features a compact design (with superconducting coils) and a low power consumption. The linac has a novel C-band high gradient structure and accelerates the ions to variable energies up to 400 MeV/u. High RF frequencies lead to power consumptions which are much lower than the ones of synchrotrons for the same ion extraction energy.

This work is part of a collaboration with the CLIC group, which is working at CERN on high-gradient electron-positron colliders.

---

[*] Corresponding author.

# Contents

## 1. Hadrontherapy and its technology

The use of light ion beams in tumor treatment was proposed when the properties of the interaction between charged particles and matter were discovered more than 60 years ago [1]. However, the developments in accelerator physics and diagnostic techniques made possible the efficient use of charged particles for tumor treatments only in recent years, with a main focus on proton beams and carbon ion beams.

     Hadrontherapy requires dedicated accelerators and sources to produce medical beams. In addition, special delivery techniques are needed to conform the delivered dose to the tumor volume. These aspects are presented in the following sections. Imaging techniques [2] and the radiobiology of ion beams [3] have been voluntarily omitted.

### 1.1 Radiotherapy with ion beams

The finite range in matter of charged particles is the main advantage offered by protons and carbon ions compared to treatments with X-rays: the energy loss curve has a peak at the end of the particle path, a few millimeters before the particle stop, so that a high dose is deposited in a localized region (the Bragg peak). The depth reached depends on the initial energy of the particle and on the irradiated material. In water, the protons and the carbon ions need respective energies of 200 MeV and 4800 MeV to reach 27 cm depth. In order to achieve the conformal delivery of the prescribed dose to the target volume and the sparing of the surrounding healthy tissues and critical structures, two different types of dose delivery systems are used: the 'broad-beam' and 'pencil-beam' techniques.

The first dose delivery system is based on a spread-out particle beam that transversally irradiates the target. This is a 'passive' system in which the beam is either scattered in two successive targets and shaped with filters, scatterers and patient-specific collimators [4], or uses beam-wobbling magnets covering the tumor cross-section with thinner scattering targets [5]. In the simplest setups, the dose cannot be tailored to the proximal end of the target volume and an undesirable dose is delivered to the adjacent normal tissue. To counteract this effect, the National Institute of Radiological Sciences (NIRS) in Japan uses the 'layer-stacking' method, in which the beam energy is changed in steps by moving a specific number of degrader plates into the beam and dynamically controlling the beam-modifying devices to adapt to the tumor shape at each energy [6].

The second more advanced dose delivery system is based on a pencil beam, which is moved point-by-point to cover the whole target volume. This is an 'active' system in which the transverse position of the beam is scanned in the tumor cross-section by two bending magnets. In parallel, the longitudinal position of the spot, corresponding to the range of incident particles, is varied either by mechanically moving absorbers or by adjusting accelerator parameters. This results in a lower secondary neutron dose to the patient body compared to a passive system. In addition, it removes the need for patient specific devices, increases the dose target volume conformality and allows the accurate modulation of the dose within the target region [7]. Two facilities have pioneered the pencil beam method for clinical use and used it to treat hundreds of patients: the Paul Scherrer Institute (PSI) with the spot scanning technique for protons [8] and the Helmholtz Center for Heavy Ion Research (GSI) with the raster scanning device for ions [9]. This technique is presently also used at the Francis H. Burr Proton Therapy Center, the MD Anderson Cancer Center, the Rhinecker Proton Therapy Center and the University of Florida Proton Therapy Institute with protons and the Heidelberg Ion-Beam Therapy Center with carbon ions and protons.

At PSI, the 8-10 mm FWHM spot is moved in relatively large steps: 75% of the spot FWHM. After irradiation of a voxel, the beam is turned off and moved to the next voxel. In the present Gantry1 system, each of the three axes is scanned with a separate device to keep the system simple, safe and reliable [10]. The first transversal and most often used motion is given by a sweeper magnet before the last 90° bending magnet. The motion in the longitudinal direction is given by placing a range shifter device in the beam, allowing to vary sequentially the proton range in water in single steps of 4.5 mm. Finally, the slowest and least frequently used motion is given by the patient table itself. In the new Gantry2, both transverse movements will be made by scanning magnets.

At GSI, a pencil beam of 4–10 mm width (FWHM) is moved in the transverse plane almost continuously in steps equal to 30 % of the spot FWHM. Indeed, the beam is not switched off, provided that the points are close enough. This requires fast scanning magnets to keep the dose applied between two points at an acceptable level. The 'painting' of successive layers is obtained by varying the extraction magnetic field of the accelerator and changing the beam energy.

A review of the results obtained by irradiating patients and of the future prospects is outside of the scope of this paper. The interested reader will find relevant information in the papers published [11, 12, 13] in the framework of the European Network for Light Ion Therapy (ENLIGHT). The conclusion was that, in the medium term, 12% (3%) of the European patients treated every year with X-rays would profit from the use of proton (carbon ion) beams. Since the average number of X-ray patients is 120000 per 10 million inhabitants, this corresponds to about 15000 proton patients and about 3500 carbon ion patients per year.

One of the main challenges in hadrontherapy is the irradiation of moving tumors [14]. To effectively accomplish this task, important technological developments are needed in [15]:
1. systems to actively scan in 3D with a pencil beam;
2. devices to detect the instantaneous position of the tumor target and produce signals to be used in feedback loops connected with the systems of point 1;
3. instruments capable of continuously monitoring the distribution of the dose in the body of the patient.

## 1.2 Accelerators

### 1.2.1 Protontherapy

Proton beams offer very good tumor dose distributions allowing dose escalation and have a radiobiological effect similar to the X-ray beams used in conventional radiotherapy [16]. Thus protons are suited for irradiation of solid tumors which being close to critical organs, cannot receive a high enough dose with X-rays with a consequent unsatisfactory control rate. This has led to the fast development of protontherapy with more than 65000 patients treated in 25 facilities worldwide, as of the end of 2009 [17].

At present, most protontherapy centers are multi-room facilities based on a normal conducting cyclotron produced by Ion Beam Applications (IBA, Belgium). The typical weight of these cyclotron magnets is around 200 tons. Cyclotrons deliver a fixed energy beam so that absorbers are used to perform tumor depth scanning. The use of absorbers needs a long and complicated Energy Selection System (ESS), typically 15 m long, and leads to neutron activation. Superconducting cyclotrons (Varian, USA) and synchrotrons (Mitsubishi, Hitachi, Japan) are also used.

Nowadays in Europe a proton treatment costs about three times more than an advanced X-ray treatment (Intensity Modulated Radiation Therapy). The future of protontherapy would be guaranteed if cheaper centers could be made commercially available. For this reason, a lot of effort is currently put in the development of proton 'single-room' facilities. A 250 MeV commercial superconducting synchrocyclotron has been designed and is currently being tested by Still River Systems (USA). The very high magnetic field of 9 Tesla allows to reduce the magnet weight to 20 tons and to mount the accelerator on a rotating gantry [18]. The Italian research Foundation TERA is working on a "Turning LInac for Protontherapy" (TULIP), a high-gradient proton linac mounted on a rotating stand [15]. In addition, two new technologies have been proposed: the Dielectric Wall Accelerator [19] and the Laser-Driven Accelerator [20]. The applicability of both concepts still needs to be proven but they could play an important role in the longer term future.

### 1.2.2 Carbon ion therapy

Carbon ion beams are proposed for the treatment of 'radioresistant' tumors, i.e. of those 10% of all solid tumors which resist to both X-ray and proton irradiation, because of their increased radiobiological effectiveness (RBE) [21]. The reason for the higher RBE is that the ionization density produced by a carbon ion traversing a cell is twenty times larger than the one of a proton having the same range, and this entails more disruptive damages to the DNA double helix. As of the end of 2009, more than 7000 patients have been treated with carbon ions, the vast majority of which have been irradiated at NIRS [17]. The results of Phase II trials are encouraging - in particular for sarcomas, liver and lung tumors - but still a lot of clinical studies have to be performed by comparing, in Phase III trials, carbon ions with protons, in order to assess for which (radioresistant) tumors, carbon ions provide a better cure.

At present, only five centers treat patients with carbon ions worldwide, and the accelerators used to bring carbon ions up to the maximum energy are all synchrotrons. These are the Heavy Ion Medical Accelerator in Chiba (HIMAC), the Heidelberg Ion Therapy Center (HIT), the Hyogo Ion Beam Medical Center (HIBMC), the Gunma University Heavy Ion Medical Center (GHMC) and the Institute of Modern Physics (IMP) in China. Carbon ion energies used for the treatment of deep-seated tumors are: 400 MeV/u at HIMAC, GHMC and IMP, 430 MeV/u at HIT and 320 MeV/u at HIBMC. In Italy, the National Center for Oncological Hadrontherapy (CNAO) featuring a synchrotron designed by the TERA Foundation will soon be operational. A synchrotron ring is typically about 20 m in diameter and can deliver beams with variable energy but with a dead time of 1-2 seconds because of the cycling magnetic field.

In this context, alternative accelerators have been proposed. The most promising ones are the Fixed Field Alternating Gradient (FFAG) accelerators [22] and the high-frequency linacs [23], boosting the energy of particles pre-accelerated either by a low energy linac or by a cyclotron. The cyclotron-linac complex has been dubbed 'cyclinac'. A high-frequency linac which accelerates protons has been prototyped and tested by TERA [24]. These accelerators will be explained in detail in the next Sections.

FFAGs can provide a high average current pulsed beam at high repetition rate. The beam can potentially be extracted at different energies without the need of absorbers. Both the variable energy and the high repetition rate are interesting for 4D multipainting of tumors. Moreover, FFAGs can be more compact than synchrotrons if the 'non-scaling' design is adopted. However, non-scaling hadron FFAGs have never been built and there are various technical difficulties related to magnet complexity, extraction techniques and the broad-band high-gradient RF cavities needed for acceleration. Overall, one can state that, for hadrontherapy, hadron linacs are closer to realization than FFAGs.

**1.3 Ion Sources for carbon ion therapy**

In carbon ion therapy, the challenge for ion sources lies in the production of fully stripped (6+) carbon ions at the right levels of current, purity and energy spread. Ion sources operating in clinical environments have to be reliable, consume low power and require little maintenance. At present, all existing and under construction carbon ion centers are synchrotrons using DC Electron Cyclotron Resonance Ion Sources (ECRIS) [25]. These produce low charge states (most often 4+) at 20-40 kV extraction voltage. The 200-400 e$\mu$A average current is captured in the synchrotron in a few milliseconds at the typical 0.5 Hz repetition rate ('multi-turn' injection). The ions are accelerated in an RFQ and then a Drift Tube Linac to 4-7 MeV/u, after which they are stripped to the 6+ charge state with close to 100% efficiency. The 4+ charge state is chosen as high as possible to reduce the dimensions of the linac injector, which is a very expensive component of the whole accelerator complex. In fact, higher charge states are more difficult to produce in sufficient quantities with an ECRIS. The 4+ charge state has the advantage of allowing the filtering of contaminant ions from Nitrogen and Oxygen, since these do not appear with charge over mass ratios of 1/3.

The Japanese centers use sources coming from the research and experience developed over the years at NIRS and more precisely, on its NIRS-ECRIS permanent magnet 10 GHz source. At HIBMC, two identical sources ECR1 and ECR2 are used, for fast switching between proton and carbon ion treatments [26]. The GHMC uses the Kei-GM, a 10 GHz permanent magnet source similar to the NIRS-Kei2 prototype [27]. At HIT and CNAO, the proton and carbon sources are permanent magnet commercial ECRIS called Supernanogans (Pantechnik, France). Finally, the IMP developed many ECRIS and currently uses the LECR3, initially designed for atomic physics [28]. Thus, ECRIS are the workhorse for hadrontherapy centers, offering reliability, commercial availability and low power consumption.

However, alternative sources have been proposed as better candidates. The most promising one is the pulsed Electron Beam Ion Source (EBIS), aiming to deliver enough current in very short pulses to achieve 'single-turn' injection in the synchrotron. This type of injection allows the synchrotron magnets aperture to be smaller, reducing the complexity and cost of the accelerator [29]. A dedicated prototype source was designed and constructed in Frankfurt, the MEDEBIS [30]. It aimed at the production of fully stripped $C^{6+}$ ions to be injected into an RFQ with much lower duty cycle ($10^{-5}$ instead of $10^{-3}$) and injected in the synchrotron without stripping. Its features were a normal conducting solenoid of 0.8 T and a short trap length of 0.2 m to reduce the pulse length to 2 μs FWHM. More recently, the new KRION2 source in operation in Russia has been proposed for injection into HIMAC [31]. The source has a superconducting solenoid of 3 T and operates in the 'Electron String' mode [32], which enhances the electron density and ion production without increasing the power consumption.

## 2. Cyclinacs

The TERA Foundation has long been active in the field of hadrontherapy. It promoted the Proton Ion Medical Machine Study [33] at CERN and designed the facility which, after the approval of the Italian Health Ministry in 2002, has been constructed by the CNAO Foundation in Pavia [34]. With the aim of addressing the technical challenges facing the future of hadrontherapy, TERA introduced in 1993 a new accelerator concept, the combination of a commercial cyclotron and a high-gradient linac, which was later dubbed 'cyclinac' [35].

### 2.1 TERA's first proposal and LIBO

The 1993 proposal was based on a 30 MeV cyclotron but later it was decided to start from the 62 MeV of the Clatterbridge cyclotron and reach 200 MeV. This linac, called LIBO for 'LInac Booster', is a side-coupled linac (SCL or also CCL – Cell Coupled Linac) operating at 3 GHz, the same frequency of all electron linacs for conventional therapy. It consisted of nine modules (each containing 52 accelerating cells) for a total length of around 13.5 m. The main advantages of this linear accelerating solution are the upgrade of the cyclotron resources already present in some hospitals, the smaller transverse beam emittances compared to the ones of cyclotrons and synchrotrons, the possibility to vary the output beam energy in only a few milliseconds and finally, the possibility of combining the spot scanning technique with multi-painting and tracking of a moving tumor.

LIBO was designed to work at a repetition rate of 200 Hz, with 5 μs RF pulses and an average accelerating gradient of 15.8 MV/m. The average beam current needed for therapy is small (nanoamperes) compared to the output current of the cyclotron, thus the $10^{-3}$ duty cycle is not a limit.

In collaboration with the Italian National Institute of Physics (INFN) and CERN, a prototype of LIBO was constructed and tested. Being the most critical part of the linac (because of the small speed of particles), the first module of the LIBO was chosen as a prototype. Full RF power tests were eventually performed at CERN, during which an accelerating gradient higher than expected (28.4 MV/m) was reached. Finally, in 2002, the prototype underwent tests with beam at the Italian Southern National Laboratories (LNS/INFN), where it successfully accelerated protons from 62 MeV to 73 MeV, demonstrating the working principle of LIBO [24].

TERA's first proposal of a full-fledged protontherapy center dates back to the year 2000, when the foundation drafted the first layout of the Institute for Diagnostics and Advanced Radiotherapy (IDRA), a unique center that combines four aspects of cancer treatment [36]:
- radioisotope production for diagnostics with PET and SPECT,
- radioisotope production for the treatment of metastases and systemic tumors through endotherapy,
- protontherapy,
- nuclear medicine research.

The rationale was to create a synergetic environment with, as its heart, a cyclinac based on the LIBO project.

## 2.2 High Gradient Linear Accelerators

### 2.2.1 Background

Since then, the TERA Foundation has pursued its research efforts. The length of a cyclinac could be reduced if high enough accelerating gradients were achieved in the linac. The main focus has thus been to achieve higher accelerating gradients, without compromising the reliability of the accelerator. However, increasing the electric field in the acceleration gaps of the linac increases the probability of electrical breakdowns in the cavity, inducing the loss of a beam pulse, and in the RF amplifiers, affecting the machine reliability. In this context, TERA is collaborating with the CLIC group at CERN, which works on high-gradient linacs for a future electron-positron collider. Although their RF structures are very different, TERA and CLIC share the same operational limits: a 250 MV/m maximum surface electric field and a maximum $10^{-7}$ breakdown rate (the number of breakdowns per pulse per meter of accelerating structure) [37, 38]. This corresponds to about one breakdown per treatment course.

Frequent cavity breakdowns have to be avoided also because, when a breakdown occurs, the vacuum is spoiled and the power reflected back to the klystron increases sharply. Experience with high-power klystrons (higher than 100 MW) has shown that the ceramic windows between the klystron and the waveguide are prone to puncture, breakdowns and 'multipactoring' [39]. However, as explained in the next Section, the powers used in the cyclinac design are one order of magnitude lower and an effective 'recirculator' (RF diode) is inserted in the waveguide circuit. Tests will soon be performed on a complete 3 GHz RF system built by A.D.A.M. S.A. (Applications of Detectors and Accelerators to Medicine, Geneva).

The RF power supply systems needed to excite an electromagnetic field in a cavity are commercialized for different frequencies. To determine the appropriate frequency at which the structure has to operate is an issue, as the frequency influences not only the beam dynamics and the cost of the machine but also, more importantly, the maximum accelerating gradient that can be achieved in the structure.

In the 1950s, Kilpatrick studied the accelerating gradients at which reliable operation could be performed as a function of the frequency. Higher accelerating gradients can be achieved for higher frequencies [40], as the semi-empirical expression given by T.J. Boyd shows (Eq. 1). In the expression, $f$ is the accelerator frequency in MHz and $E$ is the maximum accelerating gradient at which reliable operation can be obtained in MV/m.

$$f = 1.64 \cdot E^2 \cdot e^{-\frac{8.5}{E}} \qquad (1)$$

However in the last years different experiments have shown that the breakdown model used by Kilpatrick does not apply, that the maximum surface field is not the proper parameter to look at and that Kilpatrick's limitation was too conservative. Thus, higher accelerating gradients than those given by the Kilpatrick threshold can be achieved. Recently, a new quantity, the modified Poynting vector, has been found to determine the breakdown rate at frequencies larger than 10 GHz [38].

### 2.2.2 Recent TERA Research

To experimentally measure the high gradient limitation of high frequency RF cavities suitable for hadron accelerators, a 3 GHz single-cell cavity was designed, built and tested by some of the authors (R. Bonomi, A. Degiovanni, M. Garlasché and S. Verdú Andrés) together with other collaborators. The 18.9 mm-length cell was designed to be excited at 200 Hz by RF pulses of 3 μs and to reach an accelerating gradient of at least 40 MV/m, which corresponds to a 260 MV/m peak surface electric field.

To reach high electric fields, the cavity geometry was optimized by means of an RF design based on 2D and 3D simulations, respectively with Poisson Superfish [41] and HFSS [42]. The structure incorporates a cooling system designed to reduce thermally induced deformations which can detune the cavity. Mechanical stresses have been evaluated using ANSYS [43]. A drawing of the setup is shown in Fig. 1.

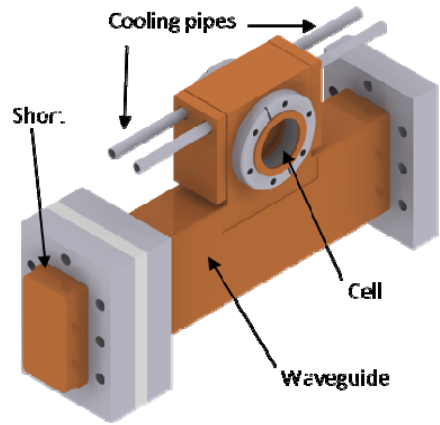

**Figure 1.** Three-dimensional drawing of the test cavity.

After construction, the cavity was tuned to the resonant frequency by deforming its nose region, with a measured Q-value within 5% of simulation results and a reflection coefficient of -27 dB. In a first high-power test performed in the CLIC Test Facility (CTF3) at CERN, the cell was operated at 50 Hz with a maximum peak input power of 1 MW. Power was provided by a 35 MW klystron delivering 5 μs pulses. A Faraday cup was connected to the cavity to measure the 'dark current'. From this signal, breakdown events were identified and the breakdown rate was estimated. The maximum electric field achieved during operation was evaluated from the power forwarded to and the power reflected by the cavity, which were monitored by a peak power meter. Contact temperature sensors were placed on the inlet and the outlet cooling pipes and at the top of the cavity, to monitor the temperature increase of the cavity.

The maximum surface electric field achieved in the cavity was above 350 MV/m, corresponding to accelerating gradients over 50 MV/m. The measured breakdown rate at these field values was around $10^{-1}$ breakdowns per pulse per meter. The preliminary maximum value of the modified Poynting vector is very close to the values achieved by high gradient accelerating structures at 12 and 30 GHz, as shown on Fig. 2.

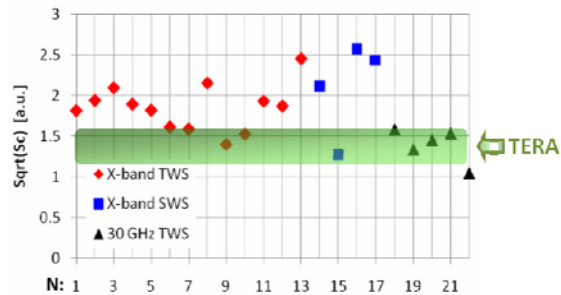

**Figure 2.** Comparison between literature [38] and test results.

The cavity suffered about 14000 breakdowns during 40 hours of operation. The cavity surface has been inspected with an optical microscope. As expected, the copper surface in the nose region (where the electric field and the Poynting vector are maximal) was damaged and some 'craters' were found. A picture of one of these areas is shown in Fig. 3.

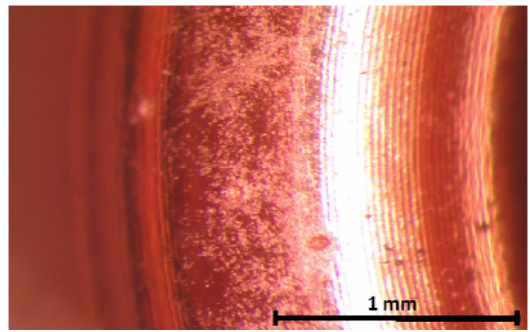

**Figure 3.** Picture of the cell cavity nose (optical microscope, 32x magnification).

These first observations are encouraging and a more precise test will be soon performed to estimate the scaling laws at 3 GHz linking the breakdown rate, the pulse length and the value of the Poynting vector.

## 3. CABOTO

CABOTO stands for 'CArbon BOoster for Therapy in Oncology' and is a direct application of the cyclinac concept to carbon ions acceleration. It combines a pulsed ion source, an (isochronous or synchro-)cyclotron and a linac. The main feature of this complex is the possibility to have intensity and energy modulated carbon ions beams well adapted to the treatment of moving organs with a 4D multi-painting spot scanning technique. Indeed, as for any cyclinac, the energy can be changed in a couple of milliseconds so that, even at a 400 Hz repetition rate, the depth reached by the spot can be adjusted at will, within the momentum acceptance of the magnetic transport line. If this acceptance is equal to ± 2 %, the range is adjustable by ± 7 %, from one spot to the next [15]. The same machine accelerates $H_2^+$ molecules, which can be stripped after acceleration to obtain proton beams. In this sense, the cyclinac is the paradigm of a dual center machine.

### 3.1 Overview of linac/cyclotron parameters

During the last years, TERA developed many designs for such a carbon cyclinac, based on different combinations of circular accelerators and linac energies and technologies. The optimal cyclotron output energy (and linac input energy) is still under investigation, but it is estimated to lie in the 120-230 MeV/u range. This value is important because it influences the size and cost of the whole machine. For instance, a high cyclotron output energy allows to reduce the length of the following linac therefore reducing the overall power consumption, but entails a bigger cyclotron magnet and reduces the active energy modulation range, as explained in the next Section.

Different technologies of circular acceleration have been considered, the most promising being the superconducting synchrocyclotron [44] and isochronous cyclotron. The first one presents some advantages in terms of power consumption and compactness. The second choice presents a less challenging design for smaller energies. Fig. 4 shows that the cyclinac solution is more compact in terms of weight than all the other circular solutions.

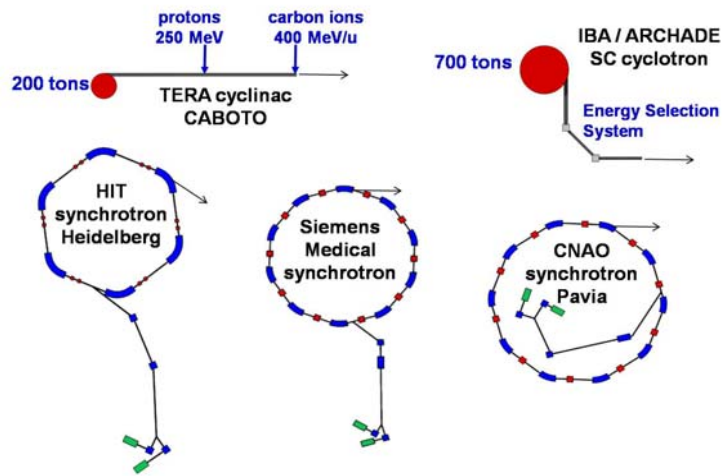

**Figure 4.** Dimensional comparison of carbon therapy accelerators [15]. The energy of the cyclotron is 120 MeV/u.

To accelerate carbon ions, one has to move towards higher linac gradients, as discussed in Section 2.2. This is because of two reasons. Firstly, fully-stripped carbon ions are made of both charged protons and neutral neutrons ($q/A=½$) so that, for the same electric field level, the energy gain per nucleon of a carbon ion is half of the energy gain of a proton ($q/A=1$). Secondly, to reach the same depth in the patient's body, carbon energies which are almost double with respect to proton energies are required.

Following the high-gradient tests at 3 GHz described in Section 2.2.2, a 5.7 GHz single cavity is under construction to bridge the gap between the experience at low frequencies and the experiments of CLIC at 12 and 30 GHz.

Waiting for the experimental results, the CABOTO design presented in Fig. 5 is based on a 120 MeV/u superconducting isochronous cyclotron, followed by a 5.7 GHz Cell Coupled Linac (CCL) which boosts the particles up to 400 MeV/u [45]. The K-480 cyclotron weighs around 150 tons, with a 2 m radius and a 2 m height. It uses superconducting coils with a 3.2 T central magnetic field.

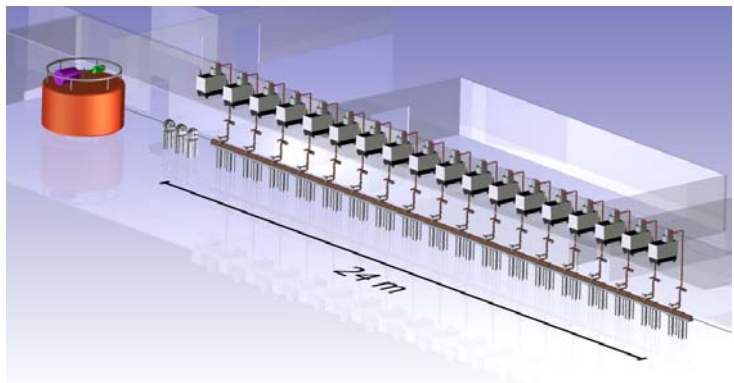

**Figure 5.** Model view of the cyclotron (120 MeV/u) and the linac (120-400 MeV/u).

The linac is composed of 18 accelerating units made of copper, each one subdivided into three tanks with a different number of cells (Fig. 6). The average length of each module is 1.3 m, for an average axial field of 40 MV/m. A series of 6 cm-long Permanent Magnet Quadrupoles (PMQs) are placed between each tank to form the 'FODO' lattice which is needed to focus the beam along the structure. Each module is powered by a 12 MW modulator-klystron system. For a 220 MW total peak RF power, the average power at 300 Hz with 2.3 µs RF pulses (filling time of 0.8 µs) is 150 kW. Taking into account the modulator-klystron efficiency, the linac consumes about 500 kW.

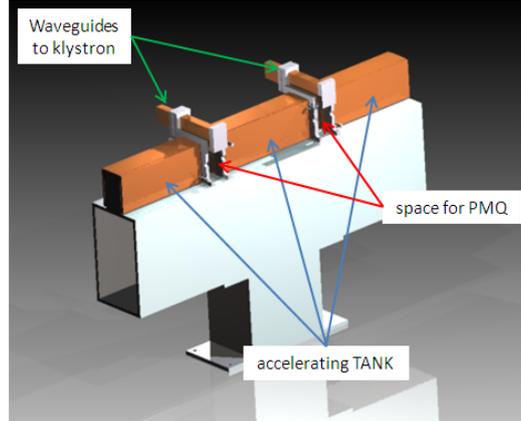

**Figure 6.** Perspective view of an accelerating unit.

### 3.2 Active Energy Modulation and Dose Delivery

The main feature of CABOTO, and of all cyclinacs, is the possibility to electronically vary the energy of the hadron beam by switching off a certain number of klystrons and by changing the amplitude and phase of the drive signal sent to the last active klystron. An illustration of this is presented in Fig. 7. This means that the range of the particles can be changed actively on a time scale of a couple of milliseconds without using passive absorbers.

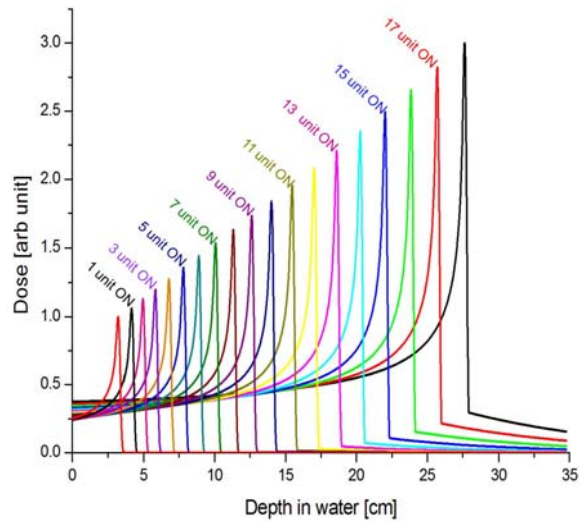

**Figure 7.** Range Variation by increasing the number of active units.

As already mentioned, the focusing of the ion beam is performed by a set of PMQs integrated in gaps located between two successive tanks, made of 17-21 accelerating cells each. The possibility to actively modulate the energy is a unique feature offered by the modularity of the linac but implies a delicate balance between the length of the tanks (i.e. the distance between successive PMQs forming the focusing FODO structure), the number of tanks powered by a single klystron and the peak power of the available klystrons. Because of the interplay between these quantities, it has been found that the most important design limitation, when considering different RF frequencies, is given by the availability of compact and cheap RF sources for the required power.

The beam dynamics has been carefully studied with the dedicated multi-particle code LINAC [46]. The purpose of the program LINAC is to simulate the motion of the charged-particle beam through a structure generated by the program DESIGN [47]. Since the magnetic gradient of the quadrupoles cannot be changed during operation, the lattice is designed in such a way that a beam arriving with energy lower than the reference one can still be transported along the linac without significant losses, even when many modules are switched off.

The maximum size of the beam profile is related to two quantities: the geometrical emittance and the Twiss Beta function $\beta x$, which depends on both the focusing strength of the quadrupoles and the RF defocusing effect of the accelerating cells. If one switches off the RF power in one accelerating tank, the corresponding RF defocusing disappears. Then, $\beta x$ depends only on the quadrupole properties and on the momentum of the particles travelling through the structure. For a traditional FODO cell (without RF defocusing), $\beta x$ has a very shallow minimum for a given phase advance. When the structure is used at energies that are lower than the maximum design value, the phase advance becomes larger than the one for which $\beta x$ has the minimum and the beam can be partially lost. It is possible to impose a limit for the phase advance, for instance by putting a limit for the FODO cell length. By imposing such a limit, one obtains the minimum value of $\beta x$ for which the beam size does not grow passing through all the tank/modules which are off, as shown in Fig. 8.

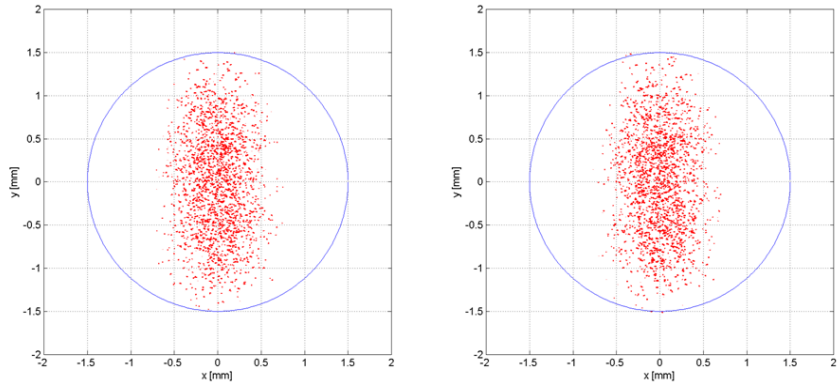

**Figure 8.** Beam transverse profile at the end of the linac with the 14[th] unit off (left) and on (right), corresponding to beam energies of 317 MeV/u and 334 MeV/u.

In order to sweep all the intermediate energies, a cyclinac requires low power klystrons in the range of 7-15 MW, in contrast with the requirements of all high frequency linacs used for research.

The fast and continuous energy variation makes the cyclinac beam more suited to the spot scanning technique with tumor multi-painting than the one produced by a cyclotron - in which the intermediate energies are obtained using passive absorbers which cause fragmentation and need a long ESS (see Fig.4) - and by a synchrotrons, in which the energy can be varied electronically, but on a time scale of the order of one second. The cyclinac beam is designed [23] to track the tumor (with an appropriate monitoring and feedback system) and paint it in a treatment session about 10 times in three dimensions so that any over-dosages and/or under-dosages can be corrected in the next painting of the same volume.

**3.3 Electron Beam Ion Source**

The cyclinac beam has particular time characteristics, as shown in Fig. 9, and requires an intense fast pulsed ion source.

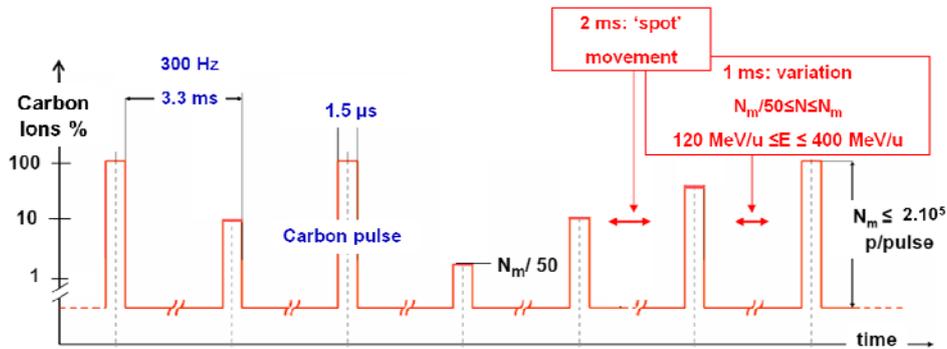

**Figure 9.** Beam structure for multipainting 4D tumor scanning.

To compute the ion output needed from the source to deliver the usual therapeutic dose rate of 2 Gray/min to a 1 L tumor, beam losses are estimated for each section of the accelerator, as shown in Table 1.

**Table 1**. Estimate of beam losses in the cyclinac.

| Source of beam loss | Estimate value |
|---|---|
| *Cyclotron injection efficiency* | 50% |
| *Cyclotron RF longitudinal capture* | 10% |
| *Cyclotron extraction efficiency* | 50% |
| *Linac transverse transmittance* | 30% |
| *Linac RF longitudinal capture* | 10 % |
| ***Overall transmittance*** | **$3.10^{-3}$** |

We see that 10% longitudinal RF capture efficiencies have to be taken into account both at injection into the cyclotron and into the linac. For this calculation, the fact that the beam extracted from the cyclotron has a microstructure of 100 MHz is irrelevant since it is seen as continuous by the 3 GHz linac.

The overall transmittance of Table 1 and the beam characteristics of Fig. 9 give the source parameters listed in Table 2. Note that the source pulse has to be longer than 1.5 μs and can have any shape (provided it is stable in time) and that $N$ is the integral particle number in 1.5 μs.

**Table 2.** Ion source requirements for cyclinacs.

|  | **Carbon Ion Therapy** | **Protontherapy** |
|---|---|---|
| *Species* | $C^{6+}$ | $H_2^+$ |
| *Repetition Rate [Hz] = R* | 300 | 100 |
| *Used Pulse Length [μs]* | 1.5 | 1.5 |
| *Particles per Pulse in 1.5 μs = N* | $\leq 1.10^8$ | $3.10^{10}$ |
| *Average Particle Intensity [s-1] = NxR* | $3.10^{10}$ | $3.10^{12}$ |

In this context, a possible candidate EBIS source for our CABOTO is the new EBIS-SC designed and built by Dreebit Gmbh [48]. This compact superconducting source is presently still under test but the expected results are very promising. Indeed, simulations show that the source could produce roughly $1.10^9$ $C^{6+}$ ions per pulse at 300 Hz repetition rate in 2-3 μs pulses [49]. Measurements with the EBIS-A (the most powerful permanent magnet EBIS from the same company) showed that the major challenge lies in the ability to produce enough carbon ions in such short ionization times (3.3 ms). The first measurements on the EBIS-SC will determine the optimal repetition rate maximizing the average particle intensity. In case this value is lower than 300 Hz, an option would be to use 2-3 sources in alternating mode, with a fast-switching system, connecting (and disconnecting) each source to (from) the cyclotron injection line.

As shown in Fig. 9, the intensity of the delivered beam needs to be modulated by factors up to 50 between one pulse and the next. The current in each beam pulse extracted from the cyclinac is at most 130 nA ($2.10^5$ carbon ions in 1.5 μs) and can be measured by pixellated ionization chambers just before the patient, as in all hadrontherapy centers. Wrong delivered doses to a specific voxel can be detected and corrected in the following paintings.

Preliminary tests made on the EBIS-A showed that the variation of the central electrode trap potential can change the number of extracted ions without modifying the pulse length, shape and ion energy. With the normal switching device used at Dreebit laboratory, the time needed to vary the electrode potential is under 1 ms, which is already fast enough for our purpose. Indeed, this would allow to achieve intensity modulation at the source level. However, this should be precisely investigated on the EBIS-SC, to determine the stability of source parameters.

Based on the working experience with the EBIS-A, we can expect impurities from $N^{7+}$ and $O^{8+}$ to be present in proportions lower than 1% [50]. This value seems acceptable in terms of therapeutic dose error. However, careful consideration has to be put into how to take this information into account in the treatment planning software.

## 4. Summary

In conclusion, pulsed ion sources and particularly, EBIS have a large potential of development for the production of fully-stripped carbon ions for hadrontherapy. They could be used for single-turn injection in synchrotrons, thus replacing the commonly used ECR ion sources, and are excellent candidate sources for the new accelerator type cyclinac. The fast and short pulsed carbon ion beam needed for spot-scanning with repainting by cyclinacs matches well the beam characteristics produced by an EBIS. In TERA's present cyclinac design, the use of the new commercial source EBIS-SC, presently under test, is foreseen. Stability, reliability and ease of use should be tested in parallel to performance, as these are crucial parameters for a medical machine.

## Acknowledgments

One of the authors (A. Garonna) would like to thank the organizers and sponsors of the International Symposium on Electron Beam Ion Sources EBIST2010, for the student grant received to attend the conference. The authors warmly thank A.D.A.M. S.A. and the Vodafone Foundation for their generous financial support of TERA research activities.